%% file: main.tex
\documentclass[journal]{IEEEtran}

\usepackage{cite}
\usepackage{amsmath,amssymb,amsfonts}
\usepackage{graphicx}
\usepackage{tikz}
\usetikzlibrary{positioning, arrows.meta, shapes.geometric, calc, fit, decorations.pathreplacing}
\usepackage{color}
\usepackage[hyphens]{url}
\usepackage{booktabs}
\usepackage{multirow}
\usepackage{algorithm}
\usepackage{algpseudocode}
\usepackage{siunitx}

\begin{document}

% === TITLE ===
\title{Grid Integration of Gigawatt-Scale Hydrogen Hubs: A Multi-Timescale Stability Analysis and Connection Requirements for Weak Grid Environments}

% === AUTHOR ===
\author{Mohamed~Shamseldein,~\IEEEmembership{Senior~Member,~IEEE}%
\thanks{M. Shamseldein is with the Department of Electrical Power and Machines, Faculty of Engineering, Ain Shams University, Cairo 11517, Egypt (e-mail: mohamed.shamseldein@eng.asu.edu.eg).}}

\maketitle

% === ABSTRACT ===
\begin{abstract}
The global transition toward green hydrogen is driving the deployment of gigawatt-scale electrolysis centers, introducing a novel, converter-dominated load class to the bulk power system. Unlike conventional industrial loads, these facilities utilize extensive power electronics interfaces with fast dynamics comparable to Inverter-Based Resources (IBRs). This paper presents a comprehensive grid impact assessment of large-scale hydrogen hubs, focusing on harmonic injection, voltage stability in low Short Circuit Ratio (SCR) environments, and frequency response capabilities.

Adopting a ``full-spectrum'' open-source modeling approach, the study utilizes \textbf{PandaPower} for large-scale steady-state contingency assessment; \textbf{ANDES} for electromechanical dynamic simulations to evaluate Fast Frequency Response (FFR); and \textbf{ParaEMT} for high-fidelity electromagnetic transient analysis of harmonic distortion and Low Voltage Ride-Through (LVRT). A critical finding of this study is that standard load models, including the generic PERC1 (data center) model, are insufficient for hydrogen hubs. The paper recommends specific structural modifications to the PERC1 model---specifically regarding process safety latches and restart voltage thresholds---to accurately capture the risk of prolonged plant tripping. Based on these findings, the paper proposes a set of standardized connection requirements to ensure secure integration.
\end{abstract}

% === KEYWORDS ===

% === SECTIONS ===
\input{sections/01_introduction}
\input{sections/02_modeling}
\input{sections/03_case_study_a}
\input{sections/04_case_study_b}
\input{sections/05_case_study_c}
\input{sections/06_requirements}
\input{sections/07_conclusion}

% === REFERENCES ===
\bibliographystyle{IEEEtran}
\bibliography{bibliography/references}

\end{document}

%% file: sections/01_introduction.tex
% Section 1: Introduction (Revision R1)

\section{Introduction}

The global transition to a low-carbon energy system has positioned green hydrogen as a critical enabler for decarbonizing hard-to-abate sectors such as heavy industry, long-haul transportation, and seasonal energy storage \cite{irena_electrolyzer_ffr}. This strategic importance has catalyzed an unprecedented scaling of hydrogen production capacity, with electrolyzer deployments transitioning from distributed kilowatt-scale pilots to centralized gigawatt-scale hydrogen hubs \cite{hydrogen_grid_impact}.

These gigawatt-scale facilities represent a fundamentally new load class for transmission system operators. Unlike traditional industrial loads characterized by rotating machinery with inherent inertia, modern electrolyzers interface with the grid through extensive power electronic converters---thyristor-based rectifiers or active IGBT-PWM front-ends---exhibiting fast dynamics comparable to those of Inverter-Based Resources (IBRs) \cite{epri_electrolyzer}. This similarity extends to their sensitivity to grid disturbances, potential for harmonic injection, and control interactions in weak grid environments with low Short Circuit Ratios (SCR) \cite{nerc_ibr_weak_grid}.

\subsection{Problem Statement}

The existing framework for connecting large industrial loads to transmission networks was developed primarily for synchronous machine-dominated systems. The tools and models used in grid planning studies---static power flow, simplified ZIP load representations, and standard dynamic models like PERC1---were not designed to capture the unique characteristics of converter-dominated loads. While commercial simulation platforms such as DIgSILENT PowerFactory and PSCAD offer integrated multi-domain analysis environments spanning RMS, EMT, and harmonic studies \cite{powerfactory_digsilent}, the open-source ecosystem lacks a comparable unified tool. More importantly, regardless of the simulation platform used, the standard load models available in these tools do not adequately represent hydrogen electrolyzer behavior. This creates three significant gaps in the current planning paradigm.

First, traditional N-1 contingency studies may underestimate voltage violations when a 500 MW electrolyzer load with specific power factor characteristics is added to a transmission network, particularly in areas with existing voltage regulation challenges.

Second, the PERC1 model, originally developed for data center UPS systems \cite{powerworld_perc1}, does not account for electrolyzer-specific safety interlocks that prevent automatic reconnection after voltage disturbances, potentially leading to optimistic frequency recovery predictions.

Third, standard planning tools cannot assess harmonic injection profiles, Low Voltage Ride-Through (LVRT) performance, or control interactions that become critical when electrolyzers are connected to weak grid locations \cite{weak_grid_scr}.

\subsection{Contributions}

This paper addresses these gaps through three contributions. We apply a multi-timescale simulation approach using PandaPower \cite{pandapower_thurner}, ANDES \cite{andes_cui}, and ParaEMT \cite{paraemt_nrel} to analyze hydrogen hub integration across timescales from microseconds to minutes. All three analysis tiers---steady-state, electromechanical, and electromagnetic transient---are performed on the IEEE 39-bus (New England) system with a unified hub location at Bus~20, ensuring consistent comparison across timescales. While multi-tool methodologies have been applied to data center integration studies, this work extends the approach to address the unique characteristics of electrolyzer loads. We propose modifications to the PERC1 load model, introducing a safety latch mechanism and adjusted restart voltage thresholds to accurately represent hydrogen electrolyzer behavior during and after grid disturbances. Based on simulation results, we develop recommended connection requirements covering active power control, reactive power capability, power quality standards per IEEE 519 \cite{ieee519} and IEEE 2800 \cite{ieee2800}, and mandatory simulation model provisions. We note that the specific thresholds for ramping rates, ACE deviations, and SCR sensitivity derived in this study are tied to the test systems used; practitioners should recalibrate these values for their specific grid conditions, as discussed in Section~\ref{sec:requirements}.

%% file: sections/02_modeling.tex
% Section 2: Technical Characteristics and Modeling (Revision R1)

\section{Technical Characteristics and Modeling}

This section presents the technical foundation for modeling large-scale hydrogen electrolyzers, including electrolyzer technologies, power electronics interfaces, the open-source simulation toolchain, and the proposed modifications to the PERC1 dynamic load model.

\subsection{Electrolyzer Technologies}

Two primary electrolyzer technologies dominate the market for grid-scale hydrogen production: Proton Exchange Membrane (PEM) and Alkaline electrolyzers \cite{pem_vs_alkaline}. Table~\ref{tab:electrolyzer_comparison} summarizes their key electrical characteristics relevant to grid integration.

\begin{table}[htbp]
\centering
\caption{Comparison of Electrolyzer Technologies}
\label{tab:electrolyzer_comparison}
\begin{tabular}{lcc}
\toprule
\textbf{Parameter} & \textbf{PEM} & \textbf{Alkaline} \\
\midrule
Current Density & $>2$ A/cm$^2$ & $0.2-0.5$ A/cm$^2$ \\
Cell Voltage & $1.5-2.0$ V & $1.8-2.2$ V \\
Ramp Rate & 100\%/s & 10-20\%/s \\
Operating Range & 0-100\% & 20-100\% \\
Response Time & Seconds & Minutes \\
Power Consumption & $51-53$ kWh/kg H$_2$ & $50-55$ kWh/kg H$_2$ \\
\bottomrule
\end{tabular}
\end{table}

The fast ramping capability of PEM electrolyzers makes them particularly suitable for providing Fast Frequency Response (FFR) services, while the slower dynamics of alkaline systems may limit their participation in such ancillary markets.

\subsection{Power Electronics Interface}

Grid-scale electrolyzers require AC-DC conversion through power electronics. Two dominant topologies exist \cite{electrolyzer_pe_interface}.

The 12-pulse thyristor rectifier uses two 6-pulse bridges with a 30$^\circ$ phase shift, cancelling the 5th and 7th harmonics while producing characteristic harmonics at orders $12k \pm 1$ (11th, 13th, 23rd, 25th). The firing angle $\alpha$ controls the DC output voltage according to
\begin{equation}
    V_{dc} = \frac{3\sqrt{2}}{\pi} V_{LL} \cos(\alpha)
\end{equation}
where $V_{LL}$ is the line-to-line AC voltage.

The IGBT-PWM active front end uses Pulse Width Modulation to achieve near-unity power factor and low Total Harmonic Distortion (THD). Harmonics appear at switching frequency sidebands ($m \cdot f_{sw} \pm n \cdot f_0$), typically well above the 50th harmonic order.

\subsection{Open-Source Toolchain}

We employ three complementary open-source tools to cover the full spectrum of grid integration phenomena, as illustrated in Fig.~\ref{fig:toolchain}. PandaPower \cite{pandapower_thurner} provides steady-state power flow and contingency analysis. ANDES \cite{andes_cui} performs electromechanical dynamic simulation for frequency stability studies. ParaEMT \cite{paraemt_nrel} enables electromagnetic transient analysis for harmonic and control interaction assessment.

We acknowledge that commercial platforms such as DIgSILENT PowerFactory \cite{powerfactory_digsilent} and PSCAD provide integrated environments that cover multiple analysis domains---including RMS, EMT, and harmonic studies---within a single scalable package. The open-source toolchain adopted here was chosen to ensure full reproducibility of results and to demonstrate that the proposed modeling methodology can be implemented without commercial license constraints. The analysis framework and Modified PERC1 model proposed in this paper are equally applicable within commercial tools.

% Include TikZ figures
\input{figures/tikz_figures}

Fig.~\ref{fig:h2_architecture} illustrates the typical grid interface architecture for a gigawatt-scale hydrogen hub, showing both rectifier topologies and the integration of FFR controllers.

Table~\ref{tab:toolchain} summarizes the phenomena addressed by each tool.

\begin{table}[htbp]
\centering
\caption{Open-Source Toolchain Capabilities}
\label{tab:toolchain}
\begin{tabular}{lccc}
\toprule
\textbf{Phenomenon} & \textbf{PandaPower} & \textbf{ANDES} & \textbf{ParaEMT} \\
\midrule
Power Flow & \checkmark & -- & -- \\
N-1 Contingency & \checkmark & -- & -- \\
Frequency Response & -- & \checkmark & -- \\
Harmonic Analysis & -- & -- & \checkmark \\
LVRT/Fault Response & -- & -- & \checkmark \\
Control Interaction & -- & -- & \checkmark \\
\bottomrule
\end{tabular}
\end{table}

\subsection{Modified PERC1 Load Model}

The PERC1 (Power Electronic Reconnecting and Ceasing) model \cite{powerworld_perc1} was developed to represent aggregated power electronic loads such as data center UPS systems. While suitable for its original purpose, it fails to capture critical aspects of electrolyzer behavior during grid disturbances.

\subsubsection{Limitations of Standard PERC1}

The standard PERC1 model assumes automatic load reconnection when voltage recovers above a threshold of approximately 0.9 p.u., a fixed time delay of about 0.5 s for reconnection, and no distinction between brief and sustained voltage dips.

For hydrogen electrolyzers, process safety interlocks prevent automatic restart after significant disturbances due to concerns about hydrogen accumulation, membrane damage, and stack potential (back-EMF) effects.

\subsubsection{Proposed Modifications}

We propose two key modifications to create a ``Modified PERC1'' for hydrogen applications.

The first modification is a safety latch. For voltage dips exceeding a critical duration $T_{critical}$ (typically 150 ms), a safety latch engages that blocks automatic reconnection. The latch holds the load offline for an extended restart interval $T_{latch}$ of 20 s after voltage recovery.
\begin{equation}
    \text{SafetyLatch} =
    \begin{cases}
        1 & \text{if } t_{dip} > T_{critical} \\
        0 & \text{otherwise}
    \end{cases}
\end{equation}

The second modification is an elevated restart threshold. The voltage recovery threshold is increased from 0.9 p.u. to 0.95 p.u. to account for electrolyzer stack back-EMF, which requires higher terminal voltage to overcome when restarting.

Algorithm~\ref{alg:perc1_modified} presents the pseudocode for the modified PERC1 logic.

\begin{algorithm}[htbp]
\caption{Modified PERC1 State Machine}
\label{alg:perc1_modified}
\begin{algorithmic}[1]
\State \textbf{Initialize:} $\text{tripped} \gets \text{False}$, $\text{latch} \gets \text{False}$, $t_{dip} \gets 0$
\For{each time step $k$}
    \If{$V_k < V_{trip}$}
        \State $t_{dip} \gets t_{dip} + \Delta t$
        \State $\text{tripped} \gets \text{True}$
        \If{$t_{dip} > T_{critical}$}
            \State $\text{latch} \gets \text{True}$ \Comment{Safety latch engaged}
        \EndIf
    \Else
        \State $t_{dip} \gets 0$
    \EndIf
    \If{$\text{tripped}$ \textbf{and} $V_k \geq V_{restart}$}
        \If{$\text{latch}$}
            \State $T_{wait} \gets T_{latch}$ \Comment{Extended restart}
        \Else
            \State $T_{wait} \gets T_{recover}$
        \EndIf
        \State $\text{tripped} \gets \text{False}$ after $T_{wait}$
    \EndIf
    \State $P_k \gets 0$ \textbf{if} $\text{tripped}$ \textbf{else} $P_0 \cdot V_k^2$
\EndFor
\end{algorithmic}
\end{algorithm}

Fig.~\ref{fig:perc1_block} provides a detailed block diagram of the Modified PERC1 load model, with the key modifications highlighted. The FracOn block (shown in red) controls the reconnection fraction based on voltage levels and dip duration. The critical changes for hydrogen applications include the elevated restart threshold and the safety latch timer that enforces an extended restart delay after severe dips.

\subsubsection{Physical Justification for Modifications}

The proposed modifications are grounded in the physical characteristics of electrolyzer systems.

Regarding hydrogen accumulation risk, during voltage disturbances the electrochemical reaction may not cease instantaneously, leading to potential hydrogen accumulation in the cell stacks. Safety systems detect this condition and require a controlled purge cycle before resuming operation, which cannot occur automatically.

Regarding stack back-EMF, when an electrolyzer stack is shut down the Nernst potential of the electrochemical cells creates a back-EMF that opposes restart. The terminal voltage must exceed this potential (typically 1.4--1.6 V per cell) to reinitiate current flow. In weak grid conditions, the POI voltage may recover to only 0.90--0.92 p.u., insufficient for restart.

Regarding membrane protection, PEM electrolyzers are particularly sensitive to differential pressure across the membrane. Rapid restart without proper pressure equalization can cause membrane damage, requiring a controlled restart sequence that standard PERC1 does not model.

%% file: figures/tikz_figures.tex
% TikZ Figures for Paper 26 (Revision R1)
% Hydrogen Hub Architecture and Modified PERC1 Model
% NOTE: Figures reordered per Reviewer 1 comment to match order of first reference in text.

% ============================================================================
% Figure 1: Multi-Timescale Simulation Toolchain (referenced first in Sec II-C)
% ============================================================================
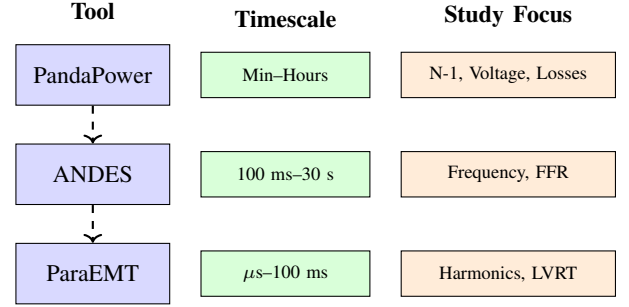
\begin{figure}[htbp]
\centering
\begin{tikzpicture}[
    node distance=0.8cm,
    tool/.style={rectangle, draw, fill=blue!15, text width=1.8cm, text centered, minimum height=0.8cm, font=\small},
    time/.style={rectangle, draw, fill=green!15, text width=2.0cm, text centered, minimum height=0.6cm, font=\scriptsize},
    study/.style={rectangle, draw, fill=orange!15, text width=2.6cm, text centered, minimum height=0.6cm, font=\scriptsize}
]

% Tools column
\node[tool] (pf) {PandaPower};
\node[tool, below=0.5cm of pf] (andes) {ANDES};
\node[tool, below=0.5cm of andes] (paraemt) {ParaEMT};

% Timescales
\node[time, right=0.4cm of pf] (t1) {Min--Hours};
\node[time, right=0.4cm of andes] (t2) {100 ms--30 s};
\node[time, right=0.4cm of paraemt] (t3) {$\mu$s--100 ms};

% Studies
\node[study, right=0.4cm of t1] (s1) {N-1, Voltage, Losses};
\node[study, right=0.4cm of t2] (s2) {Frequency, FFR};
\node[study, right=0.4cm of t3] (s3) {Harmonics, LVRT};

% Headers
\node[above=0.2cm of pf, font=\bfseries\small] {Tool};
\node[above=0.2cm of t1, font=\bfseries\small] {Timescale};
\node[above=0.2cm of s1, font=\bfseries\small] {Study Focus};

% Arrows between tools (workflow)
\draw[->, thick, dashed] (pf.south) -- (andes.north);
\draw[->, thick, dashed] (andes.south) -- (paraemt.north);

\end{tikzpicture}
\caption{Open-source simulation toolchain providing complementary analysis across timescales from steady-state planning studies to sub-cycle EMT phenomena.}
\label{fig:toolchain}
\end{figure}

% ============================================================================
% Figure 2: Hydrogen Hub Grid Interface Architecture (referenced second in Sec II-C)
% ============================================================================
\begin{figure*}[htbp]
\centering
\begin{tikzpicture}[
    node distance=1.5cm,
    block/.style={rectangle, draw, fill=blue!10, text width=2.5cm, text centered, minimum height=1.2cm, rounded corners},
    converter/.style={rectangle, draw, fill=orange!20, text width=2cm, text centered, minimum height=1cm},
    equipment/.style={rectangle, draw, fill=green!10, text width=2.5cm, text centered, minimum height=1cm},
    arrow/.style={->, >=Stealth, thick},
    bus/.style={rectangle, fill=black, minimum width=5mm, minimum height=3cm}
]

% Grid Connection
\node[block] (grid) {Transmission Grid\\(345 kV)};
\node[equipment, right=of grid] (xfmr) {Step-Down\\Transformer};
\node[bus, right=of xfmr] (poi) {};
\node[above=0.1cm of poi] {\textbf{POI}};

% Two branches - 12-pulse and PWM
\node[converter, right=1.5cm of poi, yshift=1cm] (rect1) {12-Pulse\\Rectifier};
\node[converter, right=1.5cm of poi, yshift=-1cm] (rect2) {IGBT-PWM\\AFE};

\node[equipment, right=of rect1] (stack1) {Alkaline\\Electrolyzer};
\node[equipment, right=of rect2] (stack2) {PEM\\Electrolyzer};

\node[block, right=2cm of poi, yshift=-3cm] (storage) {H$_2$ Storage\\Tank};

% Connections
\draw[arrow] (grid) -- (xfmr);
\draw[arrow] (xfmr) -- (poi);
\draw[arrow] (poi.east) -- ++(0.3,0) |- (rect1);
\draw[arrow] (poi.east) -- ++(0.3,0) |- (rect2);
\draw[arrow] (rect1) -- node[above, font=\small]{DC} (stack1);
\draw[arrow] (rect2) -- node[above, font=\small]{DC} (stack2);
\draw[arrow] (stack1.east) -- ++(0.3,0) |- (storage);
\draw[arrow] (stack2.east) -- ++(0.3,0) |- (storage);

% Labels
\node[below=0.3cm of storage] {\textbf{Hydrogen Hub (500 MW)}};

% FFR Controller (Pset label removed per Reviewer 1 comment)
\node[block, fill=red!10, below=0.5cm of poi] (ffr) {FFR\\Controller};
\draw[<->, dashed] (ffr) -- (poi);
\draw[->, dashed] (ffr.east) -- ++(0.5,0) |- (rect2);

\end{tikzpicture}
\caption{Grid interface architecture for gigawatt-scale hydrogen hub showing dual rectifier topologies (12-pulse thyristor and IGBT-PWM active front end) with FFR controller integration.}
\label{fig:h2_architecture}
\end{figure*}
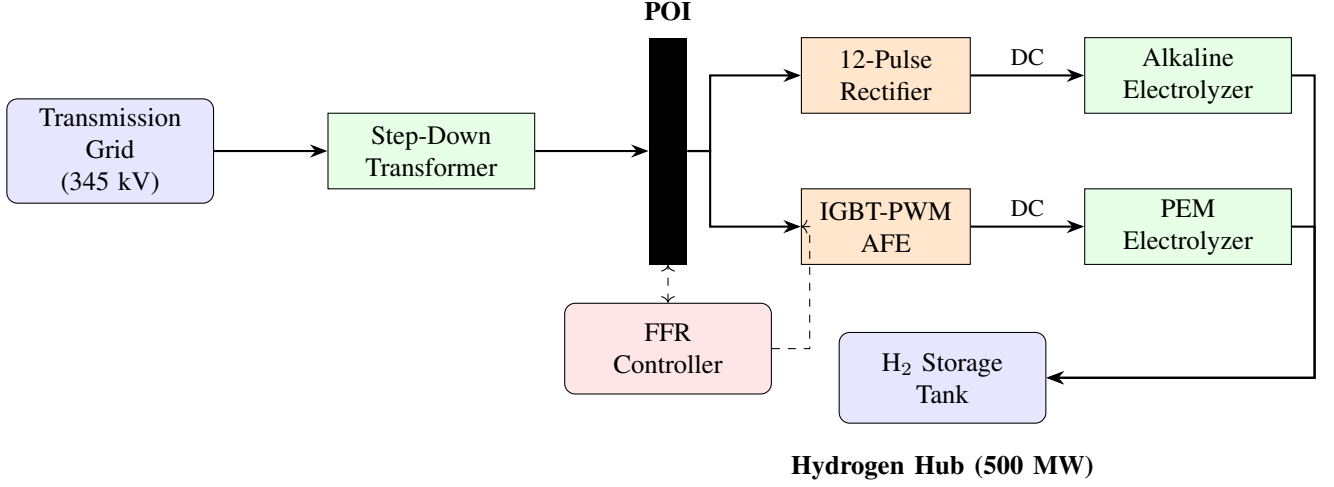

% ============================================================================
% Figure 3: Modified PERC1 Block Diagram (referenced third in Sec II-D)
% ============================================================================
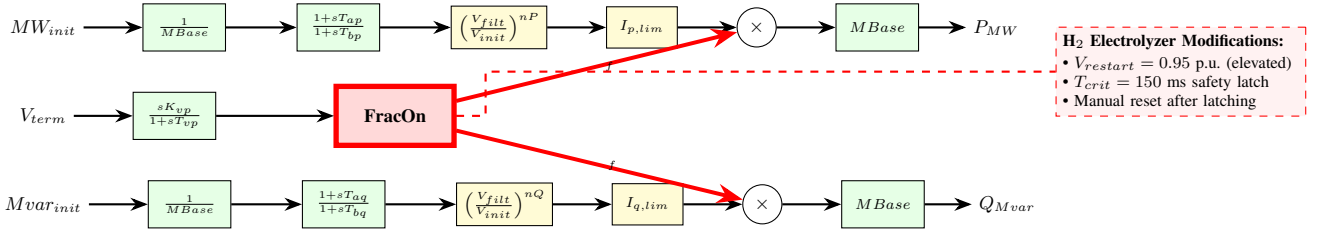
\begin{figure*}[htbp]
\centering
\begin{tikzpicture}[scale=0.78, transform shape,
    node distance=1.2cm,
    block/.style={rectangle, draw, fill=blue!10, minimum width=1.4cm, minimum height=0.8cm, font=\scriptsize},
    tf/.style={rectangle, draw, fill=green!10, minimum width=1.4cm, minimum height=0.8cm, font=\scriptsize},
    modified/.style={rectangle, draw=red, line width=2pt, fill=red!15, minimum width=2cm, minimum height=1cm, font=\small\bfseries},
    limit/.style={rectangle, draw, fill=yellow!20, minimum width=1.2cm, minimum height=0.7cm, font=\scriptsize},
    mult/.style={circle, draw, fill=white, minimum size=0.5cm, font=\small},
    arrow/.style={->, >=Stealth, thick},
    signal/.style={font=\small}
]

% === P Path (Top) ===
\node[signal] (mwinit) at (0, 1.5) {$MW_{init}$};
\node[tf, right=1cm of mwinit] (mbase1) {$\frac{1}{MBase}$};
\node[tf, right=1.2cm of mbase1] (tap) {$\frac{1+sT_{ap}}{1+sT_{bp}}$};
\node[limit, right=1.2cm of tap] (vfilt_p) {$\left(\frac{V_{filt}}{V_{init}}\right)^{nP}$};
\node[limit, right=1cm of vfilt_p] (ipmax) {$I_{p,lim}$};
\node[mult, right=1cm of ipmax] (mult_p) {$\times$};
\node[tf, right=1cm of mult_p] (mbase_out) {$MBase$};
\node[signal, right=0.8cm of mbase_out] (pmw) {$P_{MW}$};

\draw[arrow] (mwinit) -- (mbase1);
\draw[arrow] (mbase1) -- (tap);
\draw[arrow] (tap) -- (vfilt_p);
\draw[arrow] (vfilt_p) -- (ipmax);
\draw[arrow] (ipmax) -- (mult_p);
\draw[arrow] (mult_p) -- (mbase_out);
\draw[arrow] (mbase_out) -- (pmw);

% === V_term and FracOn (Middle) ===
\node[signal] (vterm) at (0, 0) {$V_{term}$};
\node[tf, right=1cm of vterm] (skvp) {$\frac{sK_{vp}}{1+sT_{vp}}$};
\node[modified, right=2cm of skvp] (fracon) {FracOn};

\draw[arrow] (vterm) -- (skvp);
\draw[arrow] (skvp) -- (fracon);
\draw[arrow, red, line width=1.5pt] (fracon) -- node[right, font=\tiny, black]{$f$} (mult_p);

% === Q Path (Bottom) ===
\node[signal] (mvarinit) at (0, -1.5) {$Mvar_{init}$};
\node[tf, right=1cm of mvarinit] (mbase2) {$\frac{1}{MBase}$};
\node[tf, right=1.2cm of mbase2] (taq) {$\frac{1+sT_{aq}}{1+sT_{bq}}$};
\node[limit, right=1.2cm of taq] (vfilt_q) {$\left(\frac{V_{filt}}{V_{init}}\right)^{nQ}$};
\node[limit, right=1cm of vfilt_q] (iqmax) {$I_{q,lim}$};
\node[mult, right=1cm of iqmax] (mult_q) {$\times$};
\node[tf, right=1cm of mult_q] (mbase_out2) {$MBase$};
\node[signal, right=0.8cm of mbase_out2] (qmvar) {$Q_{Mvar}$};

\draw[arrow] (mvarinit) -- (mbase2);
\draw[arrow] (mbase2) -- (taq);
\draw[arrow] (taq) -- (vfilt_q);
\draw[arrow] (vfilt_q) -- (iqmax);
\draw[arrow] (iqmax) -- (mult_q);
\draw[arrow] (mult_q) -- (mbase_out2);
\draw[arrow] (mbase_out2) -- (qmvar);
\draw[arrow, red, line width=1.5pt] (fracon) -- node[right, font=\tiny, black]{$f$} (mult_q);

% === Modification Callout (Right Side) ===
\node[draw=red, dashed, fill=red!5, text width=4cm, font=\footnotesize, align=left, right=0.5cm of pmw, yshift=-0.75cm] (modbox) {
\textbf{H$_2$ Electrolyzer Modifications:}\\[2pt]
\textbullet~$V_{restart} = 0.95$ p.u. (elevated)\\
\textbullet~$T_{crit} = 150$ ms safety latch\\
\textbullet~Manual reset after latching
};

% === Arrow from FracOn to callout ===
\draw[red, dashed, thick] (fracon.east) -- ++(0.5,0) |- (modbox.west);

\end{tikzpicture}
\caption{Modified PERC1 block diagram for hydrogen electrolyzers. The FracOn block (red) controls the reconnection fraction $f$ that multiplies both P and Q outputs. Key modifications include elevated restart threshold and safety latch timer.}
\label{fig:perc1_block}
\end{figure*}

%% file: sections/03_case_study_a.tex
% Section 3: Steady-State Planning Limits and N-1 Screening (Revision R1)

\section{Steady-State Planning Limits and N-1 Security Screening}

This section presents steady-state power flow and contingency screening for integrating a 500 MW hydrogen hub into the IEEE 39-bus (New England) test system using PandaPower \cite{pandapower_thurner}. The same test system is used for all three analysis tiers in this paper.

\subsection{System Configuration}

The IEEE 39-bus system was selected as the unified test system for all analyses in this study, providing 35 transmission lines, 11 transformers, and 10 generators (9 PV plus 1 slack) \cite{ieee39bus}. A 500 MW hydrogen electrolyzer was connected to Bus~20, the highest-load PQ bus in the system (existing load: 680 MW). The electrolyzer was modeled with a power factor of 0.95 lagging, consistent with typical rectifier characteristics. We note that this benchmark system, while widely used for planning studies, does not represent any specific real-world grid. The numerical thresholds derived here (e.g., voltage sensitivity coefficients, compensation sizing) are specific to this test case; application to actual power systems requires recalibration with system-specific parameters and, ideally, validation against field measurements.

\subsection{Steady-State Planning Limits}

To establish the impact of the hydrogen hub on the voltage profile, a comparative power flow analysis was performed for the original IEEE 39-bus system (without hub) and the modified system (with 500 MW hub at Bus~20). Table~\ref{tab:case_a_comparison} summarizes the results.

\begin{table}[htbp]
\centering
\caption{Power Flow Comparison: With and Without Hydrogen Hub}
\label{tab:case_a_comparison}
\begin{tabular}{lcc}
\toprule
\textbf{Metric} & \textbf{Without Hub} & \textbf{With Hub} \\
\midrule
System $V_{\min}$ & 0.982 p.u. (Bus 31) & 0.965 p.u. (Bus 20) \\
Min PQ-Bus Voltage & 0.991 p.u. (Bus 20) & 0.965 p.u. (Bus 20) \\
Hub Bus Voltage (Bus 20) & 0.991 p.u. & 0.965 p.u. \\
Max Line Loading & 73.4\% & 84.8\% \\
Total Losses & 43.6 MW & 48.6 MW \\
\bottomrule
\end{tabular}
\end{table}

The comparison reveals two important effects. First, the hydrogen hub causes a significant localized voltage depression at the POI: the hub bus voltage drops from 0.991~p.u. to 0.965~p.u. ($\Delta V = 0.026$~p.u.), and Bus~20 becomes the new system-wide voltage minimum. Without the hub, the system minimum is 0.982~p.u. at Bus~31 (a generator bus with a low voltage setpoint). Second, the maximum line loading increases from 73.4\% to 84.8\%, reflecting the additional power transfer to the hub location. Total system losses increase by 11\%.

\begin{figure}[htbp]
\centering
\includegraphics[width=\columnwidth]{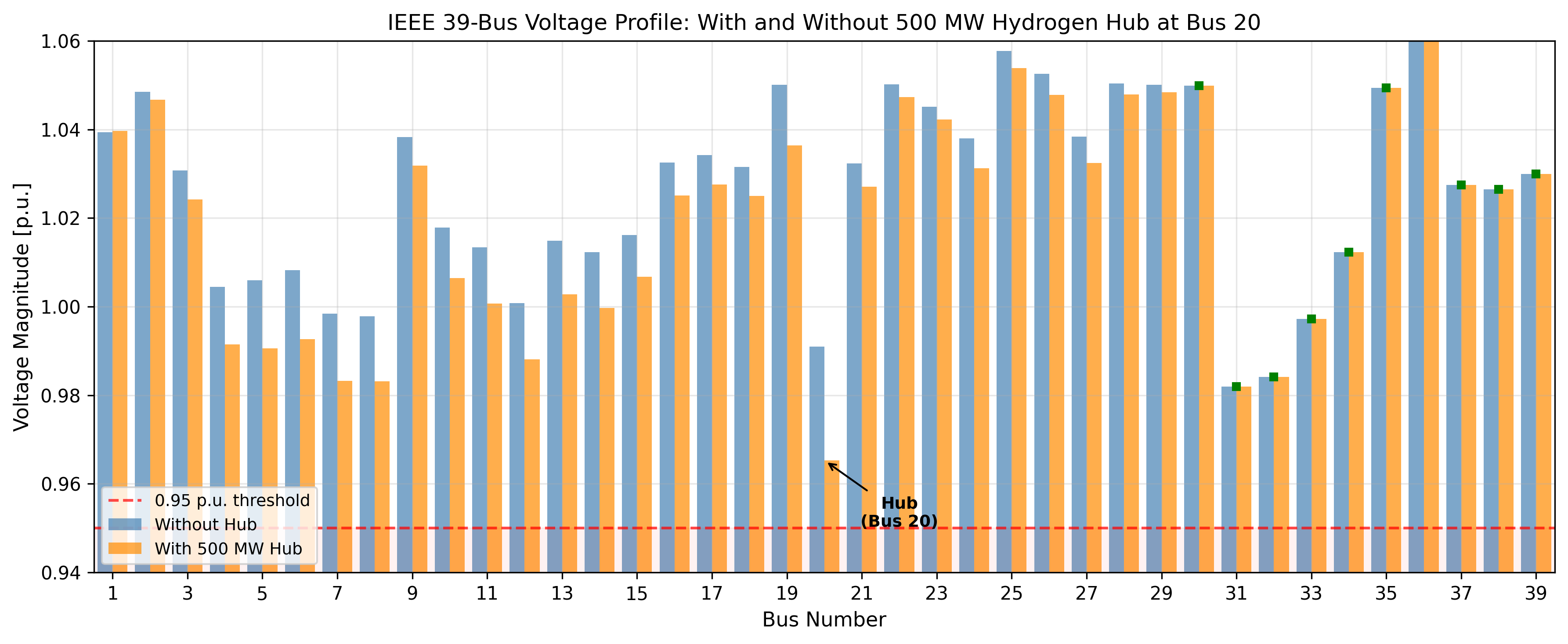}
\caption{Bus voltage profile with 500 MW hydrogen hub at Bus 20. Shaded region indicates voltages below the 0.95 p.u. planning threshold.}
\label{fig:voltage_profile}
\end{figure}

Fig.~\ref{fig:voltage_profile} illustrates the voltage distribution across all 39 buses with the hydrogen hub connected.

\subsubsection{Reactive Power Compensation}

The hub bus voltage of 0.965~p.u. is above the 0.95~p.u. planning threshold under N-0 (normal) conditions, so no reactive compensation is required at the POI for the 500 MW base case. However, the N-1 analysis in Section~\ref{sec:n1screening} reveals significant voltage violations that necessitate compensation.

\subsubsection{Voltage Sensitivity Analysis}

To understand the voltage sensitivity to electrolyzer loading, we performed a parametric study varying the hydrogen hub power from 0 to 1000 MW in 100 MW increments. Above 300 MW, Bus~20 becomes the system voltage minimum, so we focus on the POI voltage (Bus~20) as the relevant metric. The relationship between hub power $P_{hub}$ and POI voltage can be approximated as
\begin{equation}
    V_{POI} \approx V_{POI,0} - k_v \cdot P_{hub}
    \label{eq:voltage_sensitivity}
\end{equation}
where $V_{POI,0} = 0.993$~p.u. is the POI voltage without the hub and $k_v = 5.9 \times 10^{-5}$~p.u./MW is the voltage sensitivity coefficient derived from regression analysis over the 0--800 MW range.

For the 500 MW hub, this relationship predicts $V_{POI} = 0.993 - 0.030 = 0.963$~p.u., consistent with the power flow result of 0.965~p.u. Maintaining the POI voltage above 0.95~p.u. limits the hub capacity to approximately 730 MW without reactive power compensation. For larger hubs, reactive support at the POI is necessary.

\subsection{N-1 Security Screening}
\label{sec:n1screening}

An automated N-1 contingency sweep was performed on all transmission lines and transformers (46 total elements) using Algorithm~\ref{alg:contingency}.

\begin{algorithm}[htbp]
\caption{N-1 Contingency Sweep}
\label{alg:contingency}
\begin{algorithmic}[1]
\State \textbf{Input} Network $\mathcal{N}$, Lines $\mathcal{L}$, Transformers $\mathcal{T}$
\State \textbf{Output} Violation counts $(n_V, n_T)$
\For{each element $e \in \mathcal{L} \cup \mathcal{T}$}
    \State Set $e$ out of service
    \State Solve power flow using Newton-Raphson
    \If{converged}
        \State $V_{min} \gets \min_{b \in \mathcal{B}} V_b$
        \State $V_{max} \gets \max_{b \in \mathcal{B}} V_b$
        \State $L_{max} \gets \max_{e' \in \mathcal{L} \cup \mathcal{T}} \text{Loading}_{e'}$
    \EndIf
    \State Restore $e$ to service
\EndFor
\State Flag violations if $V_{min} < 0.95$ or $V_{max} > 1.05$ or $L_{max} > 100\%$
\end{algorithmic}
\end{algorithm}

\subsubsection{Violation Summary}

The N-1 sweep was performed for the base case (without hub), the hub case, and the hub with a 50 Mvar STATCOM at Bus~20. In the base case, only 2 of 46 contingencies produce voltage violations below 0.95~p.u. (worst: 0.937~p.u. at Bus~15). With the hub, 8 of 46 contingencies produce violations, with the worst case deteriorating to 0.836~p.u. at the hub bus (Bus~20). This represents a 0.101~p.u. worsening of the N-1 voltage floor attributable to the hub. No thermal violations were observed in either case.

Table~\ref{tab:case_a_scenarios} compares the three planning scenarios. A 50 Mvar STATCOM at the POI improves the worst hub-bus voltage under N-1 from 0.836 to 0.849~p.u., a modest improvement that indicates additional compensation or network reinforcement is required for full N-1 compliance.

\begin{table}[htbp]
\centering
\caption{Scenario Comparison for N-1 Screening}
\label{tab:case_a_scenarios}
\small
\begin{tabular}{lccc}
\toprule
\textbf{Scenario} & \textbf{N-1 $V_{\min}$ [p.u.]} & \textbf{N-1 Violations} \\
\midrule
Base (no hub) & 0.937 (Bus 15) & 2 of 46 \\
With hub & 0.836 (Bus 20) & 8 of 46 \\
Hub + 50 Mvar STATCOM & 0.849 (Bus 20) & 8 of 46 \\
\bottomrule
\end{tabular}
\end{table}

\begin{figure}[htbp]
\centering
\includegraphics[width=\columnwidth]{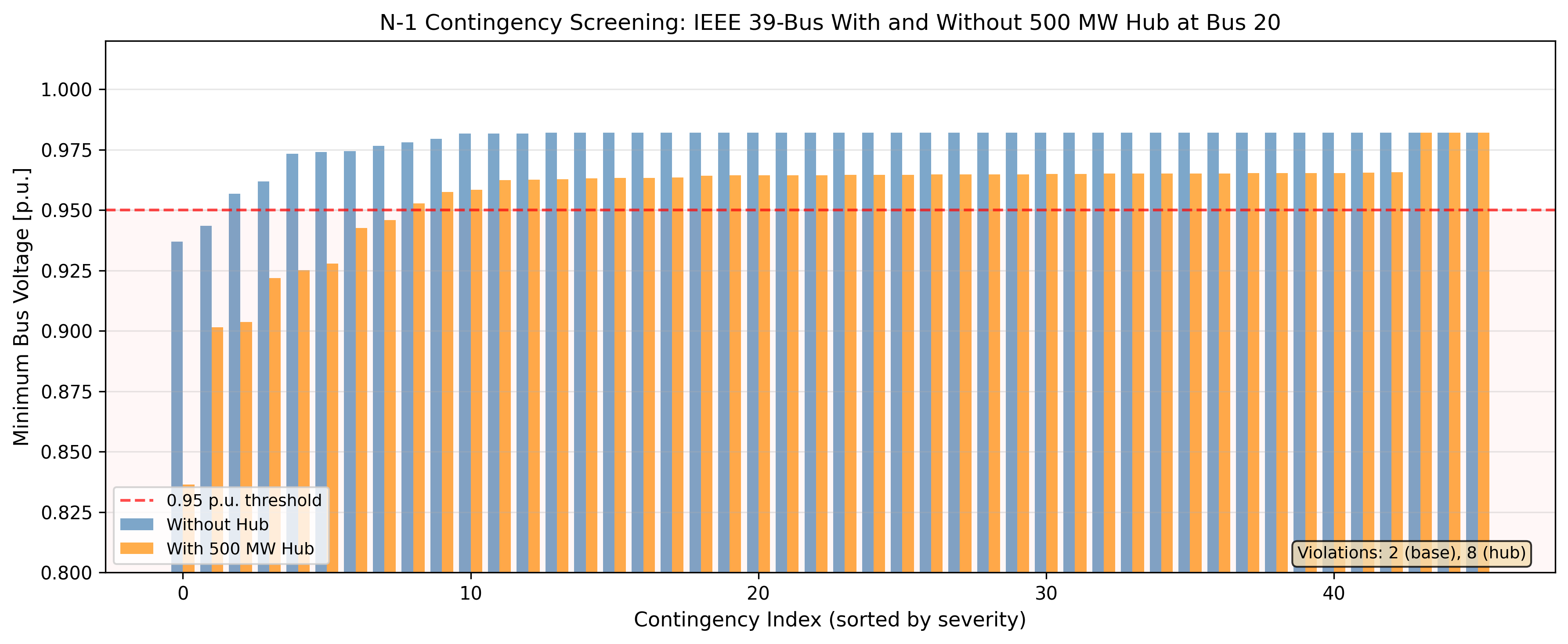}
\caption{N-1 contingency minimum voltage distribution for the IEEE 39-bus system with 500 MW hub at Bus~20. The dashed line indicates the 0.95 p.u. planning threshold.}
\label{fig:contingency_local}
\end{figure}

\subsection{Discussion}

The comparative analysis (with and without hub) confirms that the hydrogen hub causes significant voltage depression at the POI ($\Delta V = 0.026$ p.u. at Bus~20) and substantially worsens N-1 security: the number of violations increases from 2 to 8 (of 46), and the worst-case voltage drops from 0.937 to 0.836~p.u. Unlike the pre-existing violations in the base case (limited to 2 contingencies), the hub-related violations are concentrated at the POI, confirming that the hub is the dominant driver of local voltage degradation.

Key planning implications include the following.
\begin{itemize}
    \item Preliminary voltage studies comparing pre- and post-hub POI voltages should be mandatory during site selection
    \item Reactive power capability requirements should be based on SCR at the POI, recognizing that the specific SCR thresholds are system-dependent
    \item Coordinated voltage support from adjacent generation should be considered
    \item The voltage sensitivity coefficient ($k_v = 5.9 \times 10^{-5}$ p.u./MW for this test case) provides a screening metric for rapid assessment, but must be recalibrated for each candidate connection point
\end{itemize}

%% file: sections/04_case_study_b.tex
% Section 4: Frequency Dynamics (Revision R1)

\section{Frequency Dynamics and Fast Frequency Response}

This section evaluates frequency dynamics following a large generator contingency using ANDES \cite{andes_cui}, comparing standard load models against the proposed Modified PERC1, two FFR implementations with actual load shedding, and an explicit latch-triggered voltage-dip case. The same IEEE 39-bus system used for steady-state analysis in Section~III is employed here.

\subsection{System Configuration}

The IEEE 39-bus (New England) system with validated GENROU generator and TGOV1N governor models was used for all frequency dynamics simulations \cite{ieee39bus}. A 500 MW hydrogen hub was added at Bus~20 (the highest-load PQ bus, consistent with the steady-state study) as a constant-power PQ load with a power factor of 0.95 lagging. The largest PV generator, GENROU\_9 (Bus~38, 764.8 MW), was tripped at $t = 1.0$~s as the initiating contingency.

\subsection{Scenario Comparison}

Six scenarios were simulated to isolate the effects of load model selection, FFR participation, and latch-triggered voltage dips.

The first three scenarios---ZIP baseline, Standard PERC1-like, and Modified PERC1-like---use the full 500 MW hub load throughout the simulation. For a remote generator trip (no voltage event at the hub bus), these three models produce identical frequency trajectories because the PERC1 voltage-dependent reconnection logic does not activate when the hub bus voltage remains near 1.0~p.u. The differentiation between these models manifests only during voltage disturbances, as demonstrated in the sixth (latch+fault) scenario.

The fourth scenario implements a timed Fast Frequency Response (FFR) by physically disconnecting 75\% of the hub load (375 MW) at $t = 1.02$~s (20 ms after the trip). The fifth scenario approximates a droop-based FFR by disconnecting 5\% of the hub load (25 MW) at $t = 1.1$~s (100 ms delay), representing the proportional response expected from a 5\% droop controller given the observed frequency deviation. The sixth scenario applies a 150~ms three-phase fault at the hub bus (Bus~20) starting at $t = 0.95$~s, with the hub converter latching offline at fault clearing ($t = 1.10$~s), simulating converter protection response.

\subsection{Results}

\begin{figure}[htbp]
\centering
\includegraphics[width=\columnwidth]{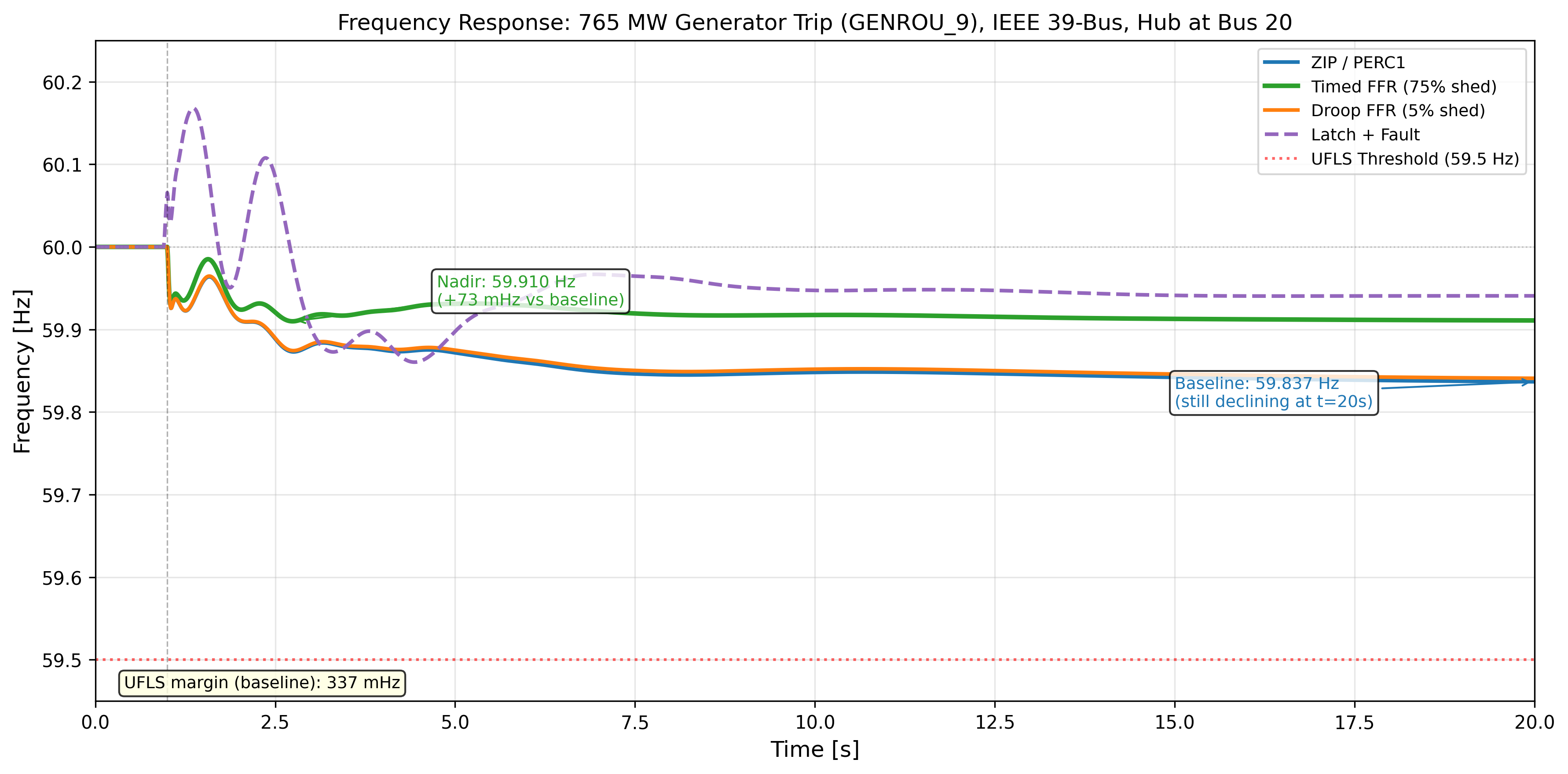}
\caption{Frequency response comparison following a 765 MW generator trip (GENROU\_9) at $t = 1.0$~s. The timed FFR case (green) shows a clear nadir improvement of 73~mHz. The latch-triggered case (purple) reaches nadir earlier and recovers to a higher final frequency because the 500 MW hub loss partially compensates for the generator loss.}
\label{fig:frequency_response}
\end{figure}

Fig.~\ref{fig:frequency_response} presents the frequency response trajectories, while Table~\ref{tab:case_b_comparison} summarizes the key metrics for all six scenarios. The first three scenarios (ZIP, Standard PERC1, Modified PERC1) produce identical traces in this event because no voltage disturbance occurs at the hub bus, so only four distinct curves appear in the figure.

\begin{table}[htbp]
\centering
\caption{Frequency Response Comparison (IEEE 39-Bus, Bus 20)}
\label{tab:case_b_comparison}
\footnotesize
\resizebox{\columnwidth}{!}{%
\begin{tabular}{lcccccc}
\toprule
\textbf{Metric} & \textbf{ZIP} & \textbf{Std} & \textbf{Mod} & \textbf{Timed FFR} & \textbf{Droop FFR} & \textbf{Latch} \\
\midrule
Nadir [Hz] & 59.837 & 59.837 & 59.837 & 59.910 & 59.841 & 59.860 \\
Nadir Time [s] & 20.0 & 20.0 & 20.0 & 2.74 & 20.0 & 4.43 \\
ROCOF [Hz/s] & 2.480 & 2.480 & 2.480 & 2.480 & 2.480 & 1.498 \\
Final $f$ [Hz] & 59.837 & 59.837 & 59.837 & 59.911 & 59.841 & 59.941 \\
\bottomrule
\end{tabular}%
}
\end{table}

Several important observations emerge from these results.

The ZIP/PERC1 baseline scenarios (1--3) show frequency declining throughout the 20~s simulation window, reaching 59.837~Hz at $t = 20$~s without full nadir arrest. The 765~MW generator loss represents approximately 12\% of total system generation on the IEEE 39-bus system, a severe contingency that challenges the governor response capacity of the remaining generators.

The timed FFR scenario demonstrates the clearest benefit of fast electrolyzer response. By shedding 375~MW of hub load 20~ms after the trip, the net power imbalance is reduced from 765~MW to approximately 390~MW. This arrests the frequency decline, producing a nadir of 59.910~Hz at $t = 2.74$~s---a 73~mHz improvement over the baseline. The frequency subsequently stabilizes near 59.911~Hz, confirming that the timed FFR effectively halts the decline.

The droop FFR scenario sheds only 25~MW (5\% of hub load), reflecting the modest proportional response expected from a 5\% droop controller given the small frequency deviation ($\Delta f \approx 0.1$~Hz at 100 ms). This produces a marginal 4~mHz improvement, confirming that droop-based FFR from a single hub provides limited benefit unless the frequency deviation is large or the droop setting is aggressive.

The latch+fault scenario produces a lower ROCOF (1.498~Hz/s vs 2.480~Hz/s) and reaches nadir at $t = 4.43$~s rather than at the simulation boundary. The hub's 500~MW loss partially compensates for the 765~MW generator loss, reducing the net imbalance to approximately 265~MW. While this results in a higher final frequency (59.941~Hz), the operational consequence---complete loss of hydrogen production---makes this scenario undesirable. We note that the magnitude of these effects is strongly dependent on the system inertia and governor response of the specific network under study; in lower-inertia systems, the nadir depression from latch-triggered load loss could be significantly more severe.

All scenarios remain above the 59.5~Hz UFLS threshold, with the baseline cases maintaining a 337~mHz margin and the timed FFR case achieving a 410~mHz margin.

\subsection{FFR Implementation Notes}

The FFR action in this study is implemented as physical load shedding using ANDES Toggle events, which disconnect a portion of the hub PQ load at specified times. The timed load shed disconnects 75\% of the hub load at $t_{trip} + 20$~ms, approximating a fast-acting supervisory control response. The droop-based FFR approximates the proportional load reduction expected from a frequency-watt controller with 5\% droop and 100~ms measurement delay. In both cases, the load reduction is permanent for the simulation duration, representing a conservative assumption; in practice, load could be gradually restored as frequency recovers.

\subsection{Discussion}

The IEEE 39-bus frequency dynamics results demonstrate that timed FFR from a large electrolyzer hub can meaningfully improve frequency nadir and arrest frequency decline following severe generator contingencies. The 73~mHz nadir improvement from shedding 75\% of a 500~MW hub represents a significant contribution to frequency stability that warrants recognition in grid connection agreements. Droop-based FFR provides more modest benefits at the contingency severity tested, suggesting that timed (deterministic) shed is preferable for large, infrequent events while droop control is better suited to continuous frequency regulation.

%% file: sections/05_case_study_c.tex
% Section 5: LVRT and Power Quality (Revision R1)

\section{LVRT and Power Quality}

This section presents EMT-level analysis using ParaEMT \cite{paraemt_nrel} for the 500 MW hydrogen hub at Bus~20 of the IEEE 39-bus system, focusing on LVRT performance and harmonic distortion at the point of interconnection (POI). The LVRT event is also repeated in ANDES to compare RMS envelopes. The same test system and hub location are used as in Sections~III and~IV.

\subsection{System Configuration}

The IEEE 39-bus (New England) system \cite{ieee39bus} was modeled in ParaEMT with a 500 MW hydrogen hub connected at Bus~20, the highest-load PQ bus. The hub size (500 MW), power factor (0.95 lagging), and connection point are consistent with the preceding steady-state and frequency analyses. Time-domain simulations were performed with a 50~$\mu$s EMT time step.

\subsection{LVRT Performance}

\begin{figure}[htbp]
\centering
\includegraphics[width=\columnwidth]{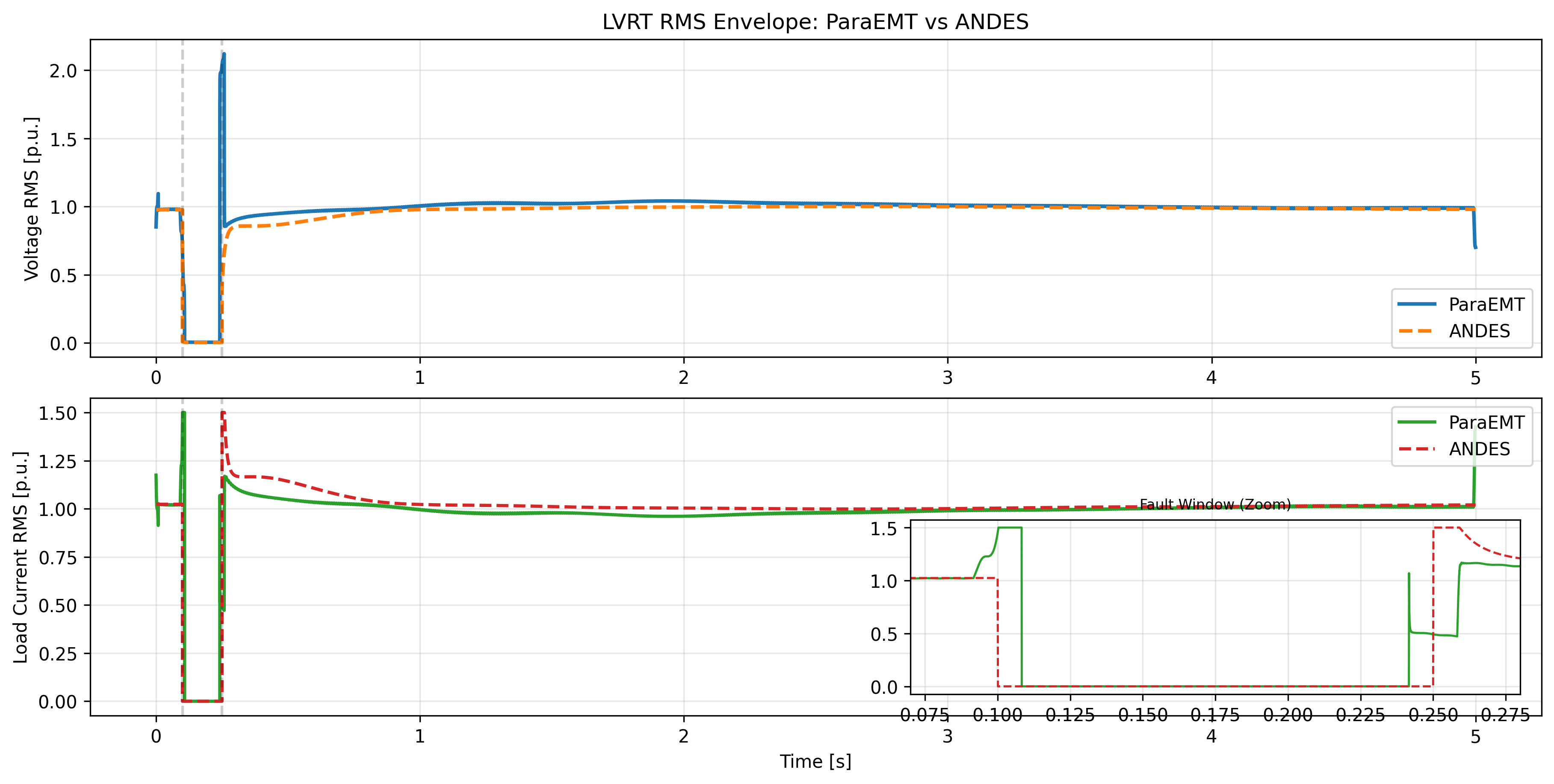}
\caption{RMS LVRT envelope comparison between ParaEMT and ANDES at Bus~20 under a three-phase bolted fault. The load current subplot shows the bounded electrolyzer current demand (I=P/V with a 1.5~p.u. limit) with a zoomed inset over the fault window.}
\label{fig:lvrt_compare}
\end{figure}

An explicit three-phase-to-ground fault was applied at the POI by adding a bolted shunt fault impedance for 150~ms, with a fault window from 0.10~s to 0.25~s. The resulting POI voltage RMS trajectory is shown in Fig.~\ref{fig:lvrt_compare}, and key metrics are summarized in Table~\ref{tab:lvrt_summary}.

\begin{table}[htbp]
\centering
\caption{LVRT Performance Summary (ParaEMT)}
\label{tab:lvrt_summary}
\begin{tabular}{lc}
\toprule
\textbf{Parameter} & \textbf{Value} \\
\midrule
Retained Voltage & 0.006 p.u. \\
Fault Duration & 150 ms \\
Pre-fault RMS Voltage & 0.981 p.u. \\
Fault Impedance & 0.0001 p.u. \\
\bottomrule
\end{tabular}
\end{table}

The POI voltage collapses to 0.006~p.u. during the bolted fault, consistent with the near-zero fault impedance. The pre-fault voltage of 0.981~p.u. reflects the voltage at Bus~20 with the 500 MW hub connected, consistent with the PandaPower steady-state result of 0.965~p.u. (the difference is attributable to the different initialization methods between the tools).

To benchmark RMS-domain fidelity, the same LVRT fault was repeated in ANDES on Bus~20. The fault is modeled as a three-phase bolted shunt with 0.0001~p.u. reactance for numerical stability. Fig.~\ref{fig:lvrt_compare} overlays the RMS voltage and load current envelopes from ParaEMT and ANDES, and Table~\ref{tab:lvrt_compare} summarizes key metrics.

Because ANDES is a phasor-domain simulator, agreement is expected only in the RMS envelope, including voltage nadir and recovery trajectory. Sub-cycle transients, phase-dependent current spikes, and harmonic behavior are outside the scope of this comparison.

Recovery time is defined as the interval from fault clearing ($t = 0.25$~s) to the first instant the bus voltage returns to its pre-fault RMS value. The load current demand $I_{load}$ is computed as $I = P/V$ on the hub MVA base with a 1.5~p.u. limit enforced to reflect converter protection; during the bolted fault ($V < 0.01$~p.u.) this demand is undefined and the converter is effectively blocked.

\begin{table}[htbp]
\centering
\caption{LVRT RMS Envelope Comparison (ParaEMT vs ANDES). Recovery is measured from fault clearing to pre-fault voltage restoration. $I_{load,\max}$ is the peak bounded converter demand ($I = P/V$, 1.5~p.u. limit) during the post-fault recovery transient.}
\label{tab:lvrt_compare}
\small
\begin{tabular}{lccc}
\toprule
\textbf{Model} & \textbf{V Nadir [p.u.]} & \textbf{Recovery [ms]} & \textbf{$I_{load,\max}$ [p.u.]} \\
\midrule
ParaEMT & 0.006 & $<$1 & 1.43 \\
ANDES & 0.004 & 723.1 & 1.50 \\
\bottomrule
\end{tabular}
\end{table}

The ParaEMT EMT-domain voltage recovers to its pre-fault RMS level within one simulation cycle after fault clearing ($<$1~ms), as the instantaneous waveform snaps back once the fault admittance is removed. The ANDES phasor-domain recovery is slower (723~ms) because it captures the post-fault electromechanical oscillations of the generator rotors. This difference is expected: EMT captures the electrical recovery, while phasor-domain captures the subsequent electromechanical settling. In the ANDES phasor-domain simulation, the slower voltage recovery causes the bounded converter demand to saturate the 1.50~p.u. limit. In the ParaEMT EMT-domain simulation, the near-instantaneous voltage snap-back limits the peak demand to 1.43~p.u.; the voltage recovers too quickly for the current to reach the 1.50~p.u. cap. Both values confirm that converter current limiting is active during the post-fault recovery transient.

\subsection{Harmonic Analysis}

Harmonic current injections were applied at the POI (Bus~20) to represent two rectifier topologies. A 12-pulse thyristor rectifier and an IGBT-PWM active front end were modeled as current source components without AC filters or converter current limits. The injected harmonic magnitudes follow typical 12-pulse characteristics (11th, 13th, 23rd, 25th) and a representative PWM spectrum with switching sidebands at orders 47, 49, 53, and 55. The resulting voltage distortion at the POI was measured via FFT. The Th\'{e}venin impedance at Bus~20, computed from the Ybus of the IEEE 39-bus system, yields a short-circuit ratio $I_{SC}/I_L \approx 19$ on the hub load base, placing this POI in the $I_{SC}/I_L < 20$ category of IEEE 519 Table~2. Table~\ref{tab:harmonic_comparison} summarizes the voltage harmonic magnitudes.

\begin{table}[htbp]
\centering
\caption{POI Voltage Harmonic Comparison (ParaEMT, Bus 20)}
\label{tab:harmonic_comparison}
\begin{tabular}{ccc}
\toprule
\textbf{Harmonic Order} & \textbf{12-Pulse [\%]} & \textbf{PWM [\%]} \\
\midrule
5th & 0.04 & 0.04 \\
7th & 0.03 & 0.03 \\
11th & 3.99 & 0.02 \\
13th & 4.03 & 0.01 \\
23rd & 4.91 & 0.01 \\
25th & 6.29 & 0.01 \\
47th & 0.00 & 3.02 \\
49th & 0.00 & 3.18 \\
53rd & 0.00 & 1.78 \\
55th & 0.00 & 1.85 \\
\midrule
\textbf{THD} & \textbf{9.79\%} & \textbf{5.09\%} \\
\bottomrule
\end{tabular}
\end{table}

\begin{figure}[htbp]
\centering
\includegraphics[width=\columnwidth]{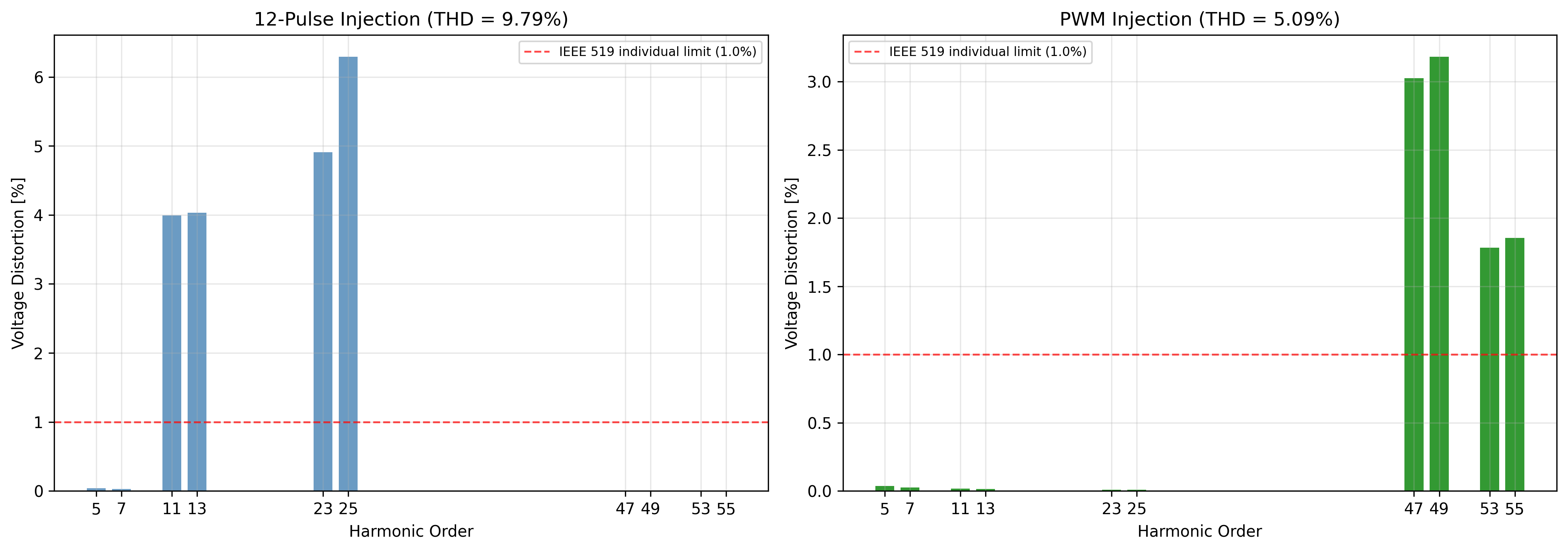}
\caption{POI voltage harmonic spectrum from ParaEMT with injected rectifier current profiles at Bus~20. The 12-pulse case shows dominant characteristic harmonics above individual IEEE 519 limits.}
\label{fig:harmonic_spectrum}
\end{figure}

The 12-pulse case exhibits dominant 11th, 13th, 23rd, and 25th harmonics, resulting in 9.79\% voltage THD at the POI. For the 345~kV POI, IEEE 519 Table~1 specifies a voltage THD limit of 1.5\% and an individual harmonic limit of 1.0\% for systems above 161~kV. Both rectifier topologies substantially exceed these limits: the 12-pulse case exceeds the 1.5\% voltage THD limit by a factor of 6.5$\times$, with four individual harmonics above the 1.0\% threshold, while the PWM case yields 5.09\% voltage THD (3.4$\times$ the limit). The PWM spectrum shifts distortion to higher-order sidebands (47th, 49th, 53rd, 55th) at 3.02\%, 3.18\%, 1.78\%, and 1.85\% respectively---all above the 1.0\% individual limit. Despite an $I_{SC}/I_L$ ratio of approximately 19 at Bus~20 (which is not an exceptionally weak bus), the large absolute magnitude of the hub's harmonic currents on a 500~MW base produces significant voltage distortion. This underscores the importance of harmonic filtering for large electrolyzer installations, particularly at high-voltage POIs where the IEEE 519 voltage distortion limits are most stringent. Fig.~\ref{fig:harmonic_spectrum} shows the spectrum comparison.

\subsection{Impedance Stability and Resonance Screening}

\begin{figure*}[htbp]
\centering
\includegraphics[width=\textwidth]{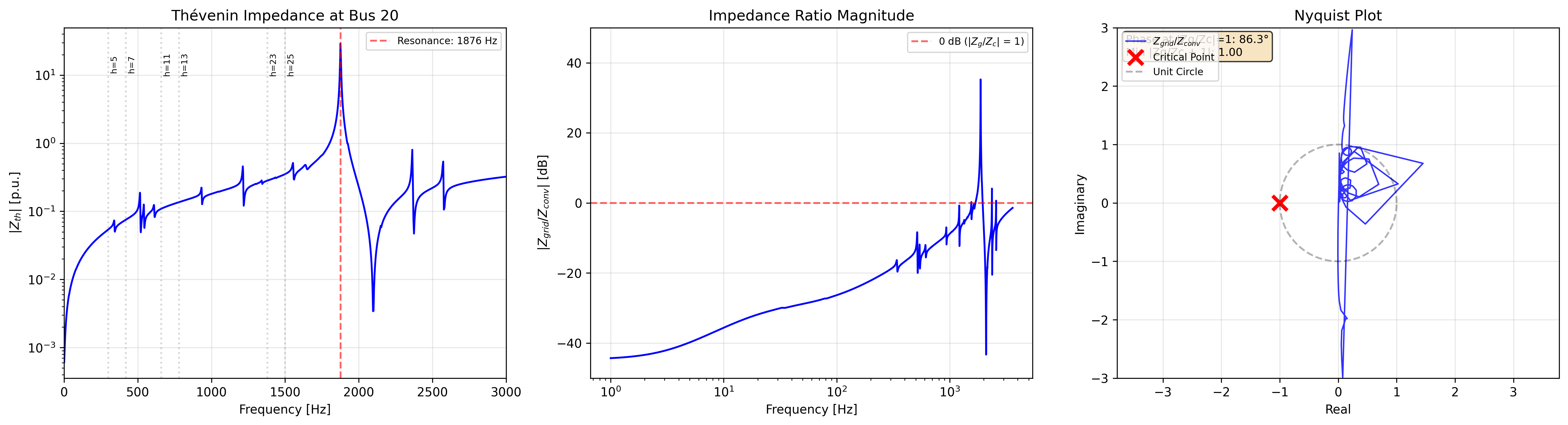}
\caption{Impedance stability screening at Bus~20, derived from the IEEE 39-bus Ybus. Left: Th\'{e}venin impedance magnitude showing a dominant parallel resonance at 1876~Hz. Center: impedance ratio magnitude. Right: Nyquist plot showing no encirclement of the critical point, with 86.3\textdegree{} phase at the unity-magnitude crossover and minimum distance 1.00.}
\label{fig:impedance_stability}
\end{figure*}

Impedance-based stability screening was performed at the POI by constructing the frequency-dependent Th\'{e}venin impedance $Z_{th}(f)$ at Bus~20 from the IEEE 39-bus Ybus matrix, with generators represented behind their subtransient reactances. The grid impedance was computed at 5000 logarithmically spaced frequency points from 1~Hz to 3600~Hz and divided by a grid-following converter impedance model incorporating PLL dynamics. Fig.~\ref{fig:impedance_stability} shows the impedance ratio and the corresponding Nyquist contour, which does not encircle the critical point. The phase at the unity-magnitude crossover is 86.3\textdegree{} and the minimum distance to the critical point is 1.00, indicating stable small-signal interaction with adequate phase margin.

The Th\'{e}venin impedance scan reveals a dominant parallel resonance at 1876~Hz (approximately the 31st harmonic order). Resonance proximity screening was conducted by computing the frequency gap between each characteristic harmonic and this resonance, along with the amplification factor (ratio of actual $|Z_{th}|$ at the harmonic frequency to the smooth inductive trend). Table~\ref{tab:resonance_screen} lists these results, with all harmonics classified as safe. The closest harmonic to the resonance is the 35th (2100~Hz) with a gap of 225~Hz; among the dominant characteristic harmonics, the 25th (1500~Hz) is closest with a gap of 376~Hz and an amplification of 1.37. Commutation risk was also checked against representative fault types, and the results indicate low risk for the tested voltage levels and durations, as shown in Table~\ref{tab:commutation_risk}. The LVRT and frequency responses show decaying oscillatory components, and no growing modes were observed over the simulated windows.

\begin{table}[htbp]
\centering
\caption{Resonance Screening Against Characteristic Harmonics}
\label{tab:resonance_screen}
\small
\begin{tabular}{cccc}
\toprule
\textbf{Harmonic} & \textbf{Freq. [Hz]} & \textbf{Gap [Hz]} & \textbf{Amp.} \\
\midrule
5th & 300 & 1576 & 1.05 \\
7th & 420 & 1456 & 1.04 \\
11th & 660 & 1216 & 0.96 \\
13th & 780 & 1096 & 1.02 \\
23rd & 1380 & 496 & 1.19 \\
25th & 1500 & 376 & 1.37 \\
35th & 2100 & 225 & 0.01 \\
37th & 2220 & 345 & 0.31 \\
\bottomrule
\end{tabular}
\end{table}

\begin{table}[htbp]
\centering
\caption{Commutation Risk Screening for Representative Faults}
\label{tab:commutation_risk}
\begin{tabular}{lccc}
\toprule
\textbf{Fault Type} & \textbf{Voltage [p.u.]} & \textbf{Duration [ms]} & \textbf{Risk} \\
\midrule
3-phase & 0.15 & 100 & Low \\
LLG & 0.35 & 150 & Low \\
SLG & 0.60 & 200 & Low \\
\bottomrule
\end{tabular}
\end{table}

\subsection{Discussion}

The ParaEMT results at Bus~20 demonstrate that even at a POI with a moderate $I_{SC}/I_L$ ratio of approximately 19, the sheer magnitude of harmonic currents from a 500~MW electrolyzer hub produces voltage THD levels (9.79\% for 12-pulse, 5.09\% for PWM) that substantially exceed the IEEE 519 voltage THD limit of 1.5\% applicable at the 345~kV POI (Table~1 of IEEE 519 for $V > 161$~kV). Harmonic filtering is therefore mandatory for large electrolyzer installations regardless of grid strength. The Ybus-derived Th\'{e}venin impedance scan confirms that the dominant parallel resonance (1876~Hz) is well separated from the characteristic harmonic frequencies of both rectifier topologies, and impedance-based Nyquist screening shows stable converter-grid interaction with adequate phase margin. The explicit LVRT fault shows that a severe short-circuit event causes complete voltage collapse at the POI, confirming that dynamic reactive support or voltage control is required for large electrolyzer hubs. Detailed EMT studies with vendor converter models and refined protection logic remain necessary for final design validation.

%% file: sections/06_requirements.tex
% Section 6: Derived Assessment Criteria (Revision R1)

\section{Derived Assessment Criteria}
\label{sec:requirements}

This section derives assessment criteria from the study results and proposes connection requirements for gigawatt-scale hydrogen hubs. The criteria are framed as evidence-based recommendations rather than fixed mandates, with thresholds tied to observed behavior in the steady-state, N-1, frequency, and EMT analyses. We emphasize that the numerical values presented here are derived from standard IEEE test systems and should be treated as starting points; system operators must adapt these thresholds to their specific grid conditions, including local short-circuit power, system inertia, and existing generation mix.

\subsection{Active Power Control}

Hydrogen hubs exceeding 100 MW should provide Fast Frequency Response (FFR) capability. The response time must not exceed 200 ms from the moment the frequency deviation exceeds the deadband, which should be set at $\pm 0.05$ Hz from nominal and be adjustable by the system operator. The droop setting should be 5\% as a default, with flexibility to adjust within the 3--7\% range depending on system conditions. The maximum FFR contribution should be capped at 50\% of rated capacity to ensure hydrogen production continuity during frequency events.

To prevent Area Control Error (ACE) issues during normal operations, ramping rate limitations are essential. The ramp-up rate should not exceed 10\%/minute under normal conditions, while ramp-down rates up to 20\%/minute may be permitted to allow faster load reduction for frequency support. Changes exceeding 100 MW should require advance notification to the system operator. We note that the appropriate ramping limits are highly dependent on the hosting power system: in interconnected continental systems with large inertia constants, higher ramp rates may be tolerable, while isolated or island systems may require more conservative limits. The ACE sensitivity to electrolyzer ramping should be assessed on a case-by-case basis using the operator's own system model.

\subsection{Reactive Power and Voltage Control}

The facility should maintain power factor capability from 0.95 leading to 0.95 lagging at the POI, with continuous variability throughout this range. The primary control mode should be voltage regulation, with power factor control available as a secondary option. The response time for reactive power changes should not exceed 1 second to reach 90\% of any step change.

During voltage disturbances, when the POI voltage falls below 0.9 p.u., the facility must prioritize reactive current injection to support voltage recovery. Active current should be limited as necessary to maintain total current at or below 1.5 p.u. The facility must maintain grid connection for all faults cleared within 150 ms, consistent with the LVRT capability demonstrated in the EMT analysis.

\subsection{Power Quality Standards}

The facility must comply with IEEE 519-2022 \cite{ieee519} at the POI for both current injection and voltage distortion. For \emph{current injection}, installations with short-circuit ratios typical of hydrogen hub locations ($I_{SC}/I_L < 20$) must limit Total Demand Distortion (TDD) to 5\%, with individual harmonics per Table~2 of the standard. For \emph{voltage distortion}, the applicable limits depend on bus voltage level: at POIs above 161~kV, IEEE 519 Table~1 specifies a voltage THD limit of 1.5\% and an individual harmonic voltage limit of 1.0\%. At the 345~kV POI studied here, the 1.5\% voltage THD limit is the binding constraint, and both rectifier topologies substantially exceed it (9.79\% for 12-pulse, 5.09\% for PWM). A pre-energization resonance study is mandatory prior to initial energization to identify potential amplification of characteristic harmonics.

IGBT-PWM active front-end rectifiers are recommended over thyristor-based designs. As demonstrated in the EMT analysis at Bus~20 ($I_{SC}/I_L \approx 19$), PWM topologies achieve 5.09\% voltage THD compared to 9.79\% for 12-pulse thyristor rectifiers---a substantial reduction, though both still exceed the 1.5\% voltage THD limit at this moderate-strength POI. Harmonic filtering is therefore mandatory regardless of rectifier topology. The specific $I_{SC}/I_L$ threshold at which PWM alone meets the voltage THD limit was not evaluated in this study and should be determined through site-specific harmonic studies for each candidate connection point.

\subsection{Impedance Stability and Resonance}

Impedance-based stability screening should be performed at the POI using the frequency-dependent Th\'{e}venin impedance derived from the network Ybus and a converter impedance model. The Nyquist plot of the impedance ratio should be checked for encirclement of the critical point. The results in this study show no Nyquist encirclement, a phase at the unity-magnitude crossover of 86.3\textdegree, and a minimum distance to the critical point of 1.00 at Bus~20 ($I_{SC}/I_L \approx 19$). Based on this evidence, a practical screening target is phase at crossover $\geq 30$\textdegree and minimum distance $|Z_g/Z_c+1| \geq 0.5$.

Resonance screening should identify the dominant parallel resonance from the Th\'{e}venin impedance magnitude scan and verify that characteristic harmonics are separated from resonance peaks by at least 60~Hz, with amplification factors (ratio of actual to smooth inductive trend) below 2.0. In this study, the dominant resonance at 1876~Hz is well separated from all characteristic harmonics (minimum gap 225~Hz at the 35th harmonic), and no amplification factor exceeds 1.37.

\subsection{Oscillation Screening}

Time-domain checks should confirm that no sustained oscillations are triggered by fault recovery, FFR actions, or control mode changes. A practical acceptance criterion is that any post-event oscillations decay without growing over a 5 s observation window.

\subsection{Simulation Model Requirements}

Facility owners must provide validated electromagnetic transient (EMT) models that include the power electronics converter topology (rectifier and DC-link), Phase-Locked Loop (PLL) representation, current and voltage controller dynamics, and LVRT and fault ride-through logic. These models are essential for assessing control interactions in weak grid environments.

For frequency stability studies, the facility must provide a dynamic load model incorporating the Modified PERC1 structure proposed in this paper. The model should include safety latch logic with a critical time threshold of 150 ms, an elevated restart voltage threshold of at least 0.95 p.u., and extended recovery time of at least 20 seconds when the safety latch is engaged. Use of standard PERC1 or ZIP models for planning studies involving hydrogen hubs is \textit{not recommended} due to the risk of underestimating load loss duration following voltage transients.

\subsection{Summary of Criteria}

Table~\ref{tab:requirements_summary} consolidates the key assessment criteria.

\begin{table}[htbp]
\centering
\caption{Assessment Criteria Summary}
\label{tab:requirements_summary}
\footnotesize
\resizebox{\columnwidth}{!}{%
\begin{tabular}{ll}
\toprule
\textbf{Category} & \textbf{Key Requirement} \\
\midrule
FFR Response Time & $\leq 200$ ms \\
Power Factor Range & 0.95 lead to 0.95 lag \\
Current TDD & $\leq 5$\% (IEEE 519 Table~2) \\
Voltage THD & $\leq 1.5$\% at $>$161~kV (IEEE 519 Table~1) \\
LVRT Duration & 150 ms at $V = 0.2$ p.u. \\
Impedance Margin & Phase at crossover $\geq 30$\textdegree{} and $|Z_g/Z_c+1| \geq 0.5$ \\
Resonance Screening & Gap $\geq 60$ Hz and amplification $< 2.0$ \\
RMS Model & Modified PERC1 mandatory \\
EMT Model & Validated model required \\
\bottomrule
\end{tabular}
}
\end{table}

%% file: sections/07_conclusion.tex
% Section 7: Conclusion (Revision R1)

\section{Conclusion}

This paper presented a comprehensive multi-timescale assessment of grid integration challenges for gigawatt-scale hydrogen hubs, employing a ``full-spectrum'' open-source simulation approach across steady-state, electromechanical, and electromagnetic transient timescales. All three analysis tiers were performed on the IEEE 39-bus (New England) system with a unified hub location at Bus~20, ensuring consistent comparison across timescales.

Comparative steady-state analysis with and without the 500 MW hydrogen hub confirmed that the hub causes significant localized voltage depression at the POI (0.026~p.u. drop at Bus~20), with the hub bus becoming the system-wide voltage minimum. N-1 contingency screening showed that the hub increases voltage violations from 2 to 8 (of 46 contingencies) and worsens the worst-case voltage from 0.937 to 0.836~p.u., underscoring the need for comprehensive voltage support planning before hydrogen hub interconnection. Frequency dynamics analysis demonstrated that timed Fast Frequency Response---shedding 75\% of hub load within 20~ms---produces a clear 73~mHz nadir improvement, arresting the frequency decline following a 765~MW generator trip. The proposed Modified PERC1 model, incorporating safety latch logic and elevated restart thresholds, more accurately represents electrolyzer behavior during grid disturbances than standard models. Harmonic analysis at the PQ bus (Bus~20) revealed that 12-pulse thyristor rectifiers produce 9.79\% voltage THD and PWM rectifiers produce 5.09\%---both substantially above the IEEE 519 voltage THD limit of 1.5\% applicable at the 345~kV POI---confirming that harmonic filtering is mandatory for electrolyzer installations at weak grid points.

We acknowledge that the numerical results are derived from a standard IEEE test system, which, while valuable for establishing methodology and relative comparisons, does not capture all characteristics of real power systems. The specific voltage sensitivities, frequency nadirs, and harmonic levels reported here should be interpreted as indicative rather than prescriptive. The connection requirements proposed in Section~\ref{sec:requirements} are intended as a starting framework that system operators should adapt based on their grid-specific conditions and, where possible, validate against field measurements from operational electrolyzer installations.

The findings support an evolution from traditional ``Load Connection Agreements'' to ``Grid Asset Agreements'' that recognize hydrogen hubs as flexible resources capable of providing stability services. Under such agreements, hydrogen facilities would receive compensation for Fast Frequency Response provision, voltage support via reactive power capability, and demand response during system stress conditions.

Future research directions include validation of the Modified PERC1 model against field measurements from operational electrolyzer facilities, extension of the analysis to real transmission system models with representative grid topologies and measured impedance data, development of a unified multi-domain test case that enables all three analysis tiers (steady-state, electromechanical, and EMT) on the same network with identical dynamic models, co-optimization of electrolyzer sizing with renewable generation and storage, and investigation of sub-synchronous control interactions in multi-converter hydrogen hub configurations.